\documentstyle[12pt,epsf]{article}

\makeatletter
\@addtoreset{equation}{section}

\makeatletter

\newcommand{\vs}[1]{\vspace*{#1}}

\newcommand{\p}{\partial}

\newcommand{\PhiP}{\Phi^{({\rm P})}}
\newcommand{\unit}{\hbox to 3.8pt{\hskip1.3pt \vrule height 7.4pt
    width .4pt \hskip.7pt \vrule height 7.85pt width .4pt \kern-2.4pt 
    \hrulefill \kern-3pt \raise 3.7pt\hbox{\char'40}}}

\begin{document}

\begin{titlepage}

\title{
\hfill\parbox{4cm}{
{\normalsize NFS-ITP-00-118}\\[-5mm]
{\normalsize\tt hep-th/0010251}
}
\\
\vspace{15mm}
Fluxons and Exact BPS Solitons 
\\
in Non-Commutative Gauge Theory
}
\author{
{}
\\
Koji {\sc Hashimoto}\thanks{{\tt koji@itp.ucsb.edu}}
\\[15pt]
{\it Institute for Theoretical Physics,}\\
{\it University of California, Santa Barbara, CA 93106}\\[7pt]
}
\date{\normalsize October, 2000}
\maketitle
\thispagestyle{empty}

\begin{abstract}
\normalsize\noindent
We show that the fluxon solution of the non-commutative gauge theory
and its variations are obtained by the soliton generation method
recently given by J.\ A.\ Harvey, P.\ Kraus and F.\ Larsen
[hep-th/0010060]. Although this method generally 
produces non-BPS solutions of equations of motion, the solutions we
obtained are BPS. We give the brane interpretation of these BPS
solutions and study their counterparts in the ordinary description by
the Seiberg-Witten map.
\end{abstract}

\end{titlepage}

%%%%%%%%%%%%%%%%%%%%%%%%%%%%%%%%%%%%%%%%%%%%%%%%%%%%%%%
%%%%%%%%%%%%%%%%%%%%%%%%%%%%%%%%%%%%%%%%%%%%%%%%%%%%%%%
%%%%%%%%%%%%%%%%%%%%%%%%%%%%%%%%%%%%%%%%%%%%%%%%%%%%%%%

\section{Introduction}

Introducing the notion of non-commutative space in string theories
have made fruitful and remarkable results in these years. Among them,
solitons in various non-commutative theories have played a central
role in understanding the physics of non-commutative theories and
certain situations of string theories. One of the applications is the 
string field theories in which introducing the non-commutativity makes
it possible to construct D-branes as solitons \cite{Dasgupta:00nt,
  Harvey:00dsns, Witten:00ntsft}. On the other hand, the 
non-commutative field theories are interesting subjects by themselves,
especially when they are realized as the low energy description of
D-branes in string theories. Solitons in these theories have
interpretation of the brane configurations, and
using the brane configuration techniques new phenomena such as  
the non-locality of the non-commutative monopoles
\cite{Hashimoto:99mdncg, Hashimoto:99bcms, Bak:99dnenbm} and the
resolution of the small instanton singularity in the moduli space of
the non-commutative instantons \cite{Nekrasov:98insst, Seiberg:99stng}
have been investigated.  

In the sequence of the above study, some exact solutions for finite
non-commutativity parameters have been constructed
\cite{Nekrasov:98insst, Furuuchi:99inrpo, Gross:00msngt, 
  Polychronakos:00ftsngt, Gross:00dsngt, Sochichiu:00ntsigf,
  bak:00esmfvbnat, 
  aganagic:00usngt, harvey:00ens, Gross:00sngt, furuuchi:00dny}. Among    
them, the transformation proposed recently by Harvey, Kraus and Larsen
(HKL) \cite{harvey:00ens} is particularly interesting, because it
transforms a trivial solution (such as a vacuum solution) into a
non-trivial classical solution of equations of motion. We summarize
their method briefly here. Let us 
consider a  
general action written in terms of various fields $\phi_i$ which are
represented by operators on the the non-commutative space. 
The equations of motion are 
\begin{eqnarray}
  \frac{\delta S}{\delta \phi_i}=0.
\end{eqnarray}
The HKL transformation is defined using the ``almost'' unitary
operator U as
\begin{eqnarray}
  \phi_i \mapsto U \phi_i U^\dagger,
\label{eq:tra}
\end{eqnarray}
where $U^\dagger U =I$, however $UU^\dagger \neq I$.
Under this transformation, the equations of motion remain intact:
\begin{eqnarray}
  \frac{\delta S}{\delta \phi_i} \mapsto 
U \frac{\delta S}{\delta \phi_i} U^\dagger=0.
\end{eqnarray}
Since to show the invariance of the equations of motion one uses only
$U^\dagger U =I$, the opposite combination
$UU^\dagger$ is not necessarily a unity. Because
$(UU^\dagger) (UU^\dagger) = U(U^\dagger U)U^\dagger =UU^\dagger$, 
the combination $UU^\dagger$ must be a projection operator.
Therefore, if we adopt some non-trivial projector $UU^\dagger$, the
transformation (\ref{eq:tra}) generates new solitons of equations of
motion. Note that we have assumed that there is no source term for the
field $\phi_i$ in the action. If the source term is present, then a
part of the equations of motion coming from the source term is not
invariant under the HKL transformation. 

In general, the HKL transformation generates non-BPS solutions.
This is because the BPS equations in non-commutative gauge theories
generally contain a constant term which becomes a source
term in the BPS equations mentioned above. 
However, in this paper, using a simple trick, we show that the BPS
fluxon solution \cite{Gross:00dsngt, Gross:00sngt} which represents a 
D-string piercing a D3-brane can be reconstructed by the HKL method. 
A virtue of this reconstruction is that we can find variations of the
fluxon solutions. These solutions have interesting brane
interpretation. Using the Seiberg-Witten map \cite{Seiberg:99stng}
which relates the 
non-commutative description to the  ordinary description, we study
the width of the fluxon and clarify the reason why there is no
corresponding solution in the ordinary description.

%%%%%%%%%%%%%%%%%%%%%%%%%%%%%%%%%%%%%%%%%%%%%%%%%%%%%%%
%%%%%%%%%%%%%%%%%%%%%%%%%%%%%%%%%%%%%%%%%%%%%%%%%%%%%%%
%%%%%%%%%%%%%%%%%%%%%%%%%%%%%%%%%%%%%%%%%%%%%%%%%%%%%%%
\section{BPS equation and solution}

Let us consider the 1+3 dimensional non-commutative gauge theory with 
a scalar field $\Phi$. We introduce non-commutativity only in the
$x^1$-$x^2$ plane: $[x^1, x^2]=i\theta$. The equations of motion in
the operator representation are written as 
\begin{eqnarray}
  [D_{\nu},[D_\nu,D_\mu]]+[\Phi,[\Phi,D_\mu]] =0,\\
{} [D_\mu,[D_{\mu}, \Phi]{}] =0.
\label{eq:eom}
\end{eqnarray}
We have defined covariant derivatives $D_\mu \equiv \p_\mu +
A_\mu$ ($\mu = 0, \cdots, 3$) where the gauge field $A_\mu$ is
anti-Hermitian. One recognizes that this equations of motion are that
of IKKT IIB Matrix model \cite{IKKT}. When written in terms of
operator language, the equations of motion can be expressed always in
the form of matrix models. This indicates a close relation between
non-commutative gauge theories and matrix models.

Assuming static configurations $\p_0=0$ and no electric field
excitation $A_0=0$, the 
above equations of motion are consistent with the following first
order BPS equation
\begin{eqnarray}
 B_{i}+ [D_i, \Phi] =0,
\label{eq:bps0}
\end{eqnarray}
where $B_i$ is the magnetic field and $i=1,2,3$. 
Taking the gauge $A_3=0$, the BPS equations become
\begin{eqnarray}
  \p_3 \Phi = -i[D_1, D_2] + \frac{1}{\theta},
\quad
\p_3 D_1 =-i [D_2, \Phi],
\quad
\p_3 D_2 =-i [\Phi,D_1].
\label{eq:bps1}
\end{eqnarray}
Defining $D\equiv (D_1 + i D_2)/\sqrt{2}$ and $\bar{D} \equiv
-D^\dagger$, we write these equations in a simple form for the latter 
convenience as  
\begin{eqnarray}
\p_3 \Phi = [D, \bar{D}] + \frac{1}{\theta},
\quad
\p_3 D = [D, \Phi].
\label{eq:bps2}
\end{eqnarray}

The HKL transformation applied to this theory is 
\begin{eqnarray}
  D \mapsto U D U^\dagger, \quad 
  \Phi \mapsto U \Phi U^\dagger.
\end{eqnarray}
So as to keep the $A_3=0$ gauge, the transformation operator $U$ has
to be independent of $x^3$. 

One notices immediately that, though this transformation keeps the
equations of motion (\ref{eq:eom}) invariant, it changes the BPS equations
(\ref{eq:bps2}). This is simply because the
constant term $1/\theta$ exists in eqs.\ (\ref{eq:bps2}). 
This constant term behaves as if it is a source term in the equation.
Generally, BPS equations are the first order equations and contain 
field strengths which are not in commutators. So the BPS equations in
the non-commutative space have the constant term which is not
invariant under the HKL transformation. 
This shows that the HKL transformation generates non-BPS solutions of
equations of motion in general. 

However, in our case, there is a certain method to obtain BPS
solutions using the HKL transformation.
Note that above BPS equations (\ref{eq:bps1}) are precisely the Nahm's
equations for the non-commutative
monopoles \cite{Bak:99dnenbm,Gross:00msngt}.  
The trick used in the papers was to redefine one of the ingredients as 
\begin{eqnarray}
  \Phi^{({\rm P})} \equiv \Phi - \frac{z}{\theta}.
\end{eqnarray}
In terms of this $\PhiP$, the BPS equations (\ref{eq:bps2}) 
do not include the constant term $1/\theta$ and they look as if they
are in the commutative space. 
\begin{eqnarray}
\p_3 \PhiP = [D, \bar{D}], \quad
\p_3 D = [D, \Phi].
\label{eq:posbps}
\end{eqnarray}
Hence we can apply the HKL transformation on these
equations without the constant term.

%%%%%%%%%%%%%%%%%%%%%%%%%%%%%%%%%%%%%%%%%%%%%%%%%%%%%%%

\begin{figure}[tbp]
\begin{center}
\leavevmode
\epsfxsize=130mm
\epsfbox{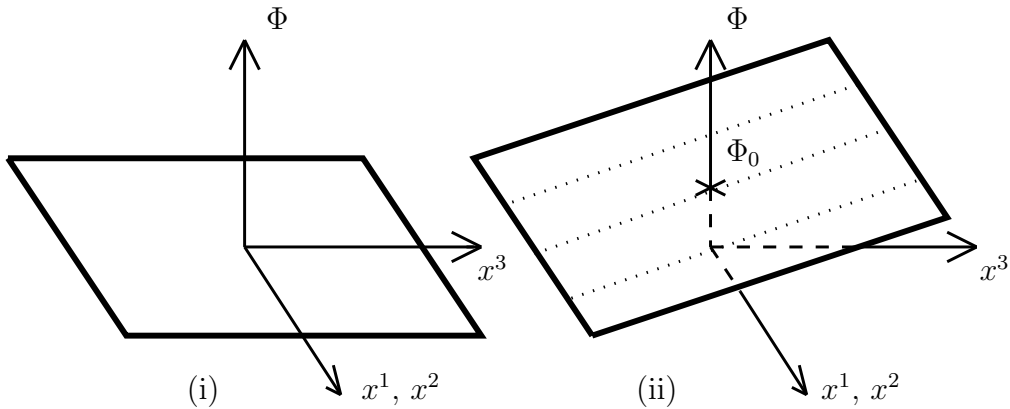}
\put(-96,90){$\Phi_0$}
\put(-96,140){$\Phi$}
\put(-270,140){$\Phi$}
\put(0,45){$x^3$}
\put(-190,45){$x^3$}
\put(-60,0){$x^1$, $x^2$}
\put(-234,0){$x^1$, $x^2$}
\put(-300,0){(i)}
\put(-130,0){(ii)}
\caption{Simple solutions to be transformed: (i) the trivial vacuum, 
(ii) the smeared D-string with no D3-brane surface. }
\label{fig:sol1}
\end{center}
\end{figure}

Before applying the transformation, let us see simple solutions to
be transformed. One of them is 
\begin{eqnarray}
\PhiP = \Phi_0 - \Phi_1 x^3, \quad D=\sqrt{\Phi_1}\; a,
\label{eq:solution}
\end{eqnarray}
where $\Phi_0$ and $\Phi_1(\geq 0)$ are real constant parameters and
$a$ is a 
creation operator: $a\equiv (x^1 + i x^2)/\sqrt{2\theta}$.
Some particular choices of these parameters exhibit interesting
solutions. First, the choice (i) $\Phi_1=1/\theta, \Phi_0=0$ provides
us with a trivial vacuum solution with $\Phi=0$, $B=0$. The second
choice (ii) $\Phi_1=0$, $\Phi_0\neq 0$ is interesting. 
In this case $\Phi=\Phi_0+z/\theta$ (see fig.\ \ref{fig:sol1}-(ii))
and we have a constant field 
strength  $B_3=-([D, \bar{D}]+\frac{1}{\theta})=-1/\theta$.
Note that the surface expressed by the scalar $\Phi$ has the constant
slope which is exactly the same slope as the one that a fluxon
solution \cite{Gross:00dsngt,Gross:00sngt} has. The fluxon solution
represents a D-string piercing the D3-brane, therefore our solution
(ii) represents a brane configuration that all the world volume is
filled with many parallel piercing D-strings. We 
cannot see the D3-brane! How this situation is possible in
the non-commutative theory and impossible in the equivalent ordinary
theory will be addressed later.

%%%%%%%%%%%%%%%%%%%%%%%%% figure %%%%%%%%%%%%%%%%%%%%%%%
%%%%%%%%%%%%%%%%%%%%%%%%% figure %%%%%%%%%%%%%%%%%%%%%%%

\begin{figure}[htdp]
\begin{center}
\leavevmode
\epsfxsize=130mm
\epsfbox{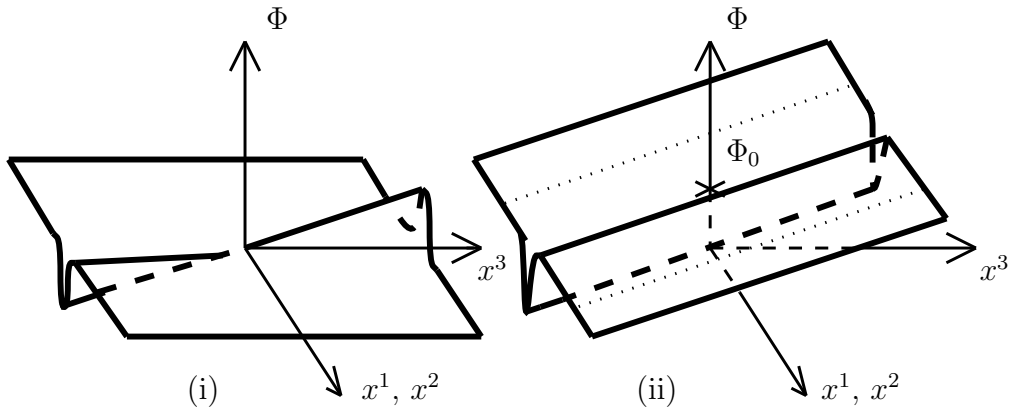}
\put(-96,90){$\Phi_0$}
\put(-96,140){$\Phi$}
\put(-270,140){$\Phi$}
\put(0,45){$x^3$}
\put(-190,45){$x^3$}
\put(-60,0){$x^1$, $x^2$}
\put(-234,0){$x^1$, $x^2$}
\put(-300,0){(i)}
\put(-130,0){(ii)}
\caption{The HKL-transformed configurations: (i) the BPS fluxon, 
(ii) a single D-string and the smeared parallel D-strings
with no D3 surface. } 
\label{fig:sol2}
\end{center}
\end{figure}

Then let us perform the HKL transformation on this solution
(\ref{eq:solution}). Taking the simplest nontrivial transformation  
$U=\sum_{n\geq 0}|n+1\rangle\langle n|$, the result for the choice (i)
is   
\begin{eqnarray}
\Phi = U \PhiP_{\rm original} U^\dagger +\frac{z}{\theta}  
=\frac{z}{\theta} P_0,\quad
B_3 = \frac{1}{\theta}P_0,
\label{eq:fluxon}
\end{eqnarray}
where $P_0$ is the projection operator onto the state
$|0\rangle$. This is precisely the BPS fluxon
\cite{Gross:00dsngt,Gross:00sngt} (see fig.\ \ref{fig:sol2}-(i)). Thus
we have reproduced the BPS  fluxon solution using the HKL
transformation. For the choice (ii), the result is 
\begin{eqnarray}
  \Phi =\Phi_0 (1-P_0) + z/\theta, \quad B_3 = -1/\theta.
\end{eqnarray}
Note that the magnetic field is not changed from the one before the
HKL transformation. From the configuration of $\Phi$ depicted in fig.\ 
\ref{fig:sol2}-(ii), we interpret this solution as the smeared many
parallel D-strings with a single D-string protruded out of them in
parallel. 
This solution is similar to the non-BPS solutions found in ref.\
\cite{aganagic:00usngt}: a certain moduli of the solution is
corresponding to the transverse separation of the object from the main
brane. In our case, the parameter $\Phi_0$ measures the separation of
a single D-string from the other smeared surface of the D-strings.
Again, we cannot see the D3-brane surface. To our best knowledge,
no similar solution has been found in the ordinary theories.

%%%%%%%%%%%%%%%%%%%%%%%%% figure %%%%%%%%%%%%%%%%%%%%%%%
%%%%%%%%%%%%%%%%%%%%%%%%% figure %%%%%%%%%%%%%%%%%%%%%%%

We can generalize the above construction easily to the non-Abelian
case. Adopting a simple solution (we are working in $U(2)$ gauge group
for simplicity)
\begin{eqnarray}
\PhiP = \Phi_1 \;z \;\unit + \Phi_0 \;\sigma_3  , \quad
D = \sqrt{\Phi_1} \;a \;\unit,
\end{eqnarray}
after the HKL transformation we obtain a generalized fluxon which
represents a D-string piercing two parallel D3-branes.  

%%%%%%%%%%%%%%%%%%%%%%%%%%%%%%%%%%%%%%%%%%%%%%%%%%%%%%%
%%%%%%%%%%%%%%%%%%%%%%%%%%%%%%%%%%%%%%%%%%%%%%%%%%%%%%%
%%%%%%%%%%%%%%%%%%%%%%%%%%%%%%%%%%%%%%%%%%%%%%%%%%%%%%%

\section{Width of the fluxon and Seiberg-Witten Map}

The fluxon solution was originally constructed by observing an
asymptotic behavior of the non-commutative $U(1)$ monopole solution
in ref.\ \cite{Gross:00dsngt}. At $x^3=+\infty$, the non-commutative
monopole solution becomes extremely simple and have the form of eq.\
(\ref{eq:fluxon}). The asymptotic value of $\Phi$ is given by the
projection operator $P_0$ which is a Gaussian of the width
$\sqrt{\theta}$ whose center is located at the origin of the
non-commutative plane ($x^1$-$x^2$ plane). Therefore the fluxon has
the width of $\sqrt{\theta}$.  

According to ref.\ \cite{Seiberg:99stng}, this non-commutative theory
has an equivalent ordinary description with the NS-NS 2-form $b$-field,
not with 
the non-commutativity. This is the ordinary Dirac-Born-Infeld theory
with the $b$-field, and in this theory a solution corresponding to the
non-commutative monopole through the Seiberg-Witten map was
constructed \cite{Hashimoto:00bbcncgt, Moriyama:00nmnm,
  Hashimoto:00sonm}. The construction of this solution is due to the
performance of the rotation in the target space. In fact, the solution 
satisfies a non-linear BPS equation, and this rotation in the target
space relates the non-linear BPS equation to the linear one, 
\begin{eqnarray}
  B_i + b_i + \p_i \Phi =0,
\end{eqnarray}
where $b_i = \epsilon_{ijk}b_{jk}/2 = b\delta_{i3}$ is the $b$-field.
The solution of this linear equation is 
\begin{eqnarray}
  \Phi = \frac{1}{r} -b x^3.
\label{eq:solor}
\end{eqnarray}
The first term (``spike'' of the BIon) shows the D-string ending on
the D3-brane  
\cite{Callan:98bdfbia,Hashimoto:99biduef}.
The second term represents a slope of the D3-brane surface.
Performing on the above linear solution (\ref{eq:solor}) the rotation
in the target space so that the D3-brane surface become horizontal,
one obtains the  
solution of the non-linear BPS equations  \cite{Hashimoto:00bbcncgt,
  Moriyama:00nmnm, Hashimoto:00sonm}. This configuration with the
horizontal D3-brane corresponds to the non-commutative monopole of
ref.\ \cite{Gross:00dsngt} through the Seiberg-Witten map.

As is seen in the following, in the ordinary description, we cannot
obtain a solution representing a D-string piercing the D3-brane. If
one wants to pierce the D3-brane, one has to add $-1/r$ to 
the above solution (\ref{eq:solor}). This $-1/r$ term represents the
D-string elongating in the $\Phi \rightarrow -\infty$ direction. 
However, this term cancels the first term in (\ref{eq:solor}) and the
whole spike vanishes. 

Another argument is as follows: The fluxon solution is obtained by
seeing the asymptotic behavior ($x^3\rightarrow \infty$) of the
non-commutative monopole. If one see the corresponding asymptotic
behavior of the solution of the ordinary side in refs.\
\cite{Moriyama:00nmnm, Hashimoto:00sonm}, it 
is easy to find that at the positive infinity of $x^3$ the solution
becomes singular. The width of the BIon is getting thinner and thinner 
in this asymptotic region. 

These arguments show that the fluxon solution in the ordinary
description does not exist. However, if one believes the validity of
the Seiberg-Witten map, the fluxon can be mapped to some configuration
in the ordinary description. Then what happens to the Seiberg-Witten
map? 

The hint for answering this question is in the solution of the choice
(ii) above (fig.\ 1-(ii)). That 
solution consists simply of many parallel D-strings, no D3-brane. Let
us pay attention to the field strength of that solution. As in ref.\
\cite{Seiberg:99stng}, the Seiberg-Witten map can be  exactly solved
for constant field strength. Particularly, when $B_3=-1/\theta$, it
was shown that there is no corresponding ordinary description: $F_{\rm 
  ordinary} = \infty$. This is the case for the choice (ii). Thus
there is no solution like this (ii) in the ordinary description.
In this sense, this solution (ii) is proper to the non-commutative
gauge theory.

Now, let us see the Seiberg-Witten map of the monopole solution
(\ref{eq:solor}). As in the same manner, when $B_3+b=0$ in the
ordinary description, the
Seiberg-Witten map becomes singular and the non-commutative
field strength diverges \cite{Seiberg:99stng}. 
At the infinity $r=\infty$ the solution (\ref{eq:solor}) satisfies 
$B_3+b>0$ (we have assumed $b>0$ for simplicity). So we cannot reach the
region $B_3+b<0$ in the non-commutative description because at the point
$B_3+b=0$ the Seiberg-Witten map becomes singular. 

The region $B+b<0$ is almost a ball with a radius $1/\sqrt{b}$ in the
world volume. At the 
surface of this ball the non-commutative field strength is diverging. 
So this radius corresponds to the size of the flux tube in the
non-commutative description. Using the essential relation $b\sim
\theta^{-1}$ \cite{Seiberg:99stng} in the Seiberg-Witten map, we
expect that the width of 
the fluxon is $\sqrt{\theta}$. This is in agreement with the explicit
solution of the fluxon (\ref{eq:fluxon}). 

This argument shows also that the non-commutative  monopole solution
of ref.\ \cite{Gross:00dsngt} corresponds to only a part of the
D3-brane surface of the solution in the ordinary description\footnote{ 
In the above argument, we have applied naively the result of the
Seiberg-Witten map of the constant field strength to the non-constant
case. However, we believe that the qualitative argument survives even
in the case of the non-constant field strength.
Another assumption used above is that we can rotate the D3-brane
surface in the target space even in the non-commutative theories, as
in the case of ordinary Dirac-Born-Infeld theory
\cite{Hashimoto:00sonm}. This assumption is not negligible: 
though the  global rotation in the target space has to be accompanied
by the diffeomorphism, in the non-commutative theories how the 
diffeomorphism acts on gauge fields is not well-defined. 
There exists the ordering ambiguities.
This difficulty is precisely the same as the one of non-Abelian
Dirac-Born-Infeld theory.
}: in 
the region $B+b<0$ of the solution (\ref{eq:solor}) there is no
non-commutative counterpart which is properly defined also at
$r=\infty$.

%%%%%%%%%%%%%%%%%%%%%%%%%%%%%%%%%%%%%%%%%%%%%%%%%%%%%%%
%%%%%%%%%%%%%%%%%%%%%%%%%%%%%%%%%%%%%%%%%%%%%%%%%%%%%%%
%%%%%%%%%%%%%%%%%%%%%%%%%%%%%%%%%%%%%%%%%%%%%%%%%%%%%%%

\section{General argument and discussion}

In sec.\ 2, we have applied the HKL transformation to the BPS
equations of the non-commutative monopoles. It needs the trick of
redefinition of the scalar field to apply the transformation.

Usually BPS equations are the first order equations, and therefore
contain terms consisting merely of field strengths $F$. 
On non-commutative space the field strength is defined using the
commutator of the derivatives which now gives a constant term. Due to
this constant term in the field strength, the BPS equations are not 
left intact under the HKL transformation, though the equations of
motion are invariant on the contrary because
they are usually the second order equations. Taking a commutation of
$F$, then the troublesome constant terms is always dropped.
Hence, usually the HKL transformation does not generate solutions
satisfying BPS equations. 

A natural question is the following: when can we apply the HKL
transformation to BPS equations?
In sec.\ 2, we have used the trick to absorb the constant term by
noting that 
the Nahm's equation is in the
same form as the BPS equations of the non-commutative monopoles
(\ref{eq:bps1}). This coincidence is because these two are related
with each other by a non-commutative analogue of the Nahm's transform
\cite{Corrigan:84}.
(The Nahm's equation can be understood as a BPS equation on D-string
worldsheet gauge theory \cite{Diaconescu:96dmne,Hashimoto:98sjwgt}.  
To deal with the non-commutative monopoles in string theory, one needs 
to put D-strings ending on parallel D3-branes, in the $b$-field
\cite{Hashimoto:99mdncg}. The above Nahm's transform is possibly
understood as T-dualities with the $b$-field in terms of string theory
(see, for example, ref.\ \cite{Hori:99dtit}), which exchange the roles
of D-strings and D3-branes.)
The Nahm's equation parameterize the moduli space of monopoles. The
trick in sec.\ 2 was used to show that the non-commutativity does not
affect the moduli space \cite{Bak:99dnenbm,Moriyama:00nmnm}. Hence,
this suggests that if the non-commutativity does not change the moduli
space then we can use the HKL transformation to obtain BPS solutions. 

This argument is consistent with the self-dual instanton solution
\cite{aganagic:00usngt} which is BPS but can be obtained by the HKL
transformation. Note that in this case the non-commutativity
$\theta^{\mu\nu}$ is also self-dual, thus
the self-dual equation describing this non-commutative instanton is
not modified. Consistently with the Nahm's transform, the ADHM
equation is not modified by the self-dual non-commutativity parameter
\cite{Seiberg:99stng}, thus the moduli space of this instanton is not
changed and includes a small instanton singularity
\cite{furuuchi:00dny}. 

This shows that the BPS solution generated by the HKL transformation
has the moduli space which is precisely the same as the ordinary version
of that. 

%%%

The above argument is applied only for the BPS equations which can be
obtained by the dimensional reduction of the self-dual instanton
equation in 4-dimensional Yang-Mills theory. 
One of the other examples of such theories is the BPS non-Abelian
vortex described by the Hitchin equation. The non-commutative version
of this BPS equation contains a constant 
term explicitly which cannot be removed by any field redefinition. 
So this BPS vortex cannot be treated by the HKL transformation.

One of the other
types of the BPS equations is on the vortex of the 1+2-dimensional
non-commutative Abelian-Higgs model \cite{Jatkar:00novnahm,
  bak:00esmfvbnat, harvey:00ens}: 
\begin{eqnarray}
  [D,\bar{\phi}]=[\bar{D},\phi]=0,\quad
B=\phi\bar{\phi}-\phi_0^2.
\label{eq:vortex}
\end{eqnarray}
Here $|\phi|=\phi_0$ is the bottom of the potential for the complex
scalar field $\phi$, and only for the special choice of the
coefficient of this potential we can achieve the saturation of the 
BPS bound with the above equations \cite{Jatkar:00novnahm}. Since the
definition of the magnetic field in the non-commutative space is
$B=[\bar{D}, D]-1/\theta$, we observe that if and only if 
$\theta=1/\phi_0^2$, the constant term in eq.\ (\ref{eq:vortex})
vanishes and we can apply the HKL transformation to obtain BPS
solutions. We shall not write the explicit solutions here because
a similar solution has already been obtained in ref.\
\cite{bak:00esmfvbnat, harvey:00ens}.

Not only the fluxons, but also the BPS non-commutative monopole
solution itself \cite{Gross:00msngt,Gross:00sngt} is possibly
generated by the HKL transformation using the trick of sec.\ 2. The
non-commutative monopole solution depends on $x^3$, thus accordingly
the soliton-generating operator  $U$ should be $x^3$-dependent. 
In sec.\ 2 we have chosen a $x^3$-independent $U$ so as to keep the
$A_3=0$ gauge. However, we can choose another prescription of the
application of the HKL transformation: assume that $A_3$ is not
transformed by any $U$. Then the remaining 
equations (\ref{eq:posbps}) is invariant under the HKL transformation
if 
\begin{eqnarray}
  [\Phi, U^\dagger\p_3 U] =
  [D, U^\dagger\p_3 U] =0.
\label{eq:x3dep}
\end{eqnarray}
In this way one can generate $x^3$-dependent BPS solutions.
However, the above constraint (\ref{eq:x3dep}) on $U(x^3)$ is turned
out to be difficult to be solved even for a specific choice of $\Phi$
and $D$ (such as the vacuum solution). We leave this issue for the
future study. 

%%%%%%%%%%%%%%%%%%%%%%%%%%%%%%%%%%%%%%%%%%%%%%%%%%%%%%
%%%%%%%%%%%%%%%%%%%%%%%%%%%%%%%%%%%%%%%%%%%%%%%%%%%%%%
%%%%%%%%%%%%%%%%%%%%%%%%%%%%%%%%%%%%%%%%%%%%%%%%%%%%%%

\vs{10mm}

\noindent
{\Large\bf Note added}

 After the completion of this paper, we became aware of the
paper \cite{Hamanaka:00oenbs} which contains the overlapping results. 

%%%%%%%%%%%%%%%%%%%%%%%%%%%%%%%%%%%%%%%%%%%%%%%%%%%%%%
%%%%%%%%%%%%%%%%%%%%%%%%%%%%%%%%%%%%%%%%%%%%%%%%%%%%%%
%%%%%%%%%%%%%%%%%%%%%%%%%%%%%%%%%%%%%%%%%%%%%%%%%%%%%%

\vs{10mm}

\noindent
{\Large\bf Acknowledgment}

We would like to thank D.\ J.\ Gross and P.\ Kraus for useful comments.
This work is supported in part by Grant-in-Aid for Scientific
Research from Ministry of Education, Science, Sports and Culture of
Japan (\#02482), and by the Japan Society for the Promotion of Science
under the Postdoctoral/Predoctoral Research Program.

%%%%%%%%%%%%%%%%%%%%%%%%%%%%%%%%%%%%%%%%%%%%%%%%%%%%%%
%%%%%%%%%%%%%%%%%%%%%%%%%%%%%%%%%%%%%%%%%%%%%%%%%%%%%%
%%%%%%%%%%%%%%%%%%%%%%%%%%%%%%%%%%%%%%%%%%%%%%%%%%%%%%

%%%%%%%%%% References %%%%%%%%%%%%%%%%%%%%%%%%%
\newcommand{\J}[4]{{\sl #1} {\bf #2} (#3) #4}
\newcommand{\andJ}[3]{{\bf #1} (#2) #3}
\newcommand{\AP}{Ann.\ Phys.\ (N.Y.)}
\newcommand{\MPL}{Mod.\ Phys.\ Lett.}
\newcommand{\NP}{Nucl.\ Phys.}
\newcommand{\PL}{Phys.\ Lett.}
\newcommand{\PR}{Phys.\ Rev.}
\newcommand{\PRL}{Phys.\ Rev.\ Lett.}
\newcommand{\PTP}{Prog.\ Theor.\ Phys.}
\newcommand{\hep}[1]{{\tt hep-th/{#1}}}
\newcommand{\hepp}[1]{{\tt hep-ph/{#1}}}
%%%%%%%%%%%%%%%%%%%%%%%%%%%%%%%%%%%%%%%%%%%%%%%

\end{document}